\documentclass[lettersize,journal]{IEEEtran}
\usepackage{amsmath,amsfonts}
\usepackage{algorithmic}
\usepackage{array}
\usepackage[caption=false,font=normalsize,labelfont=sf,textfont=sf]{subfig}
\usepackage{textcomp}
\usepackage{stfloats}
\usepackage{url}
\usepackage{verbatim}
\usepackage{graphicx}

\usepackage[utf8]{inputenc} % allow utf-8 input
\usepackage[T1]{fontenc}    % use 8-bit T1 fonts
\usepackage{hyperref}       % hyperlinks
\usepackage{url}            % simple URL typesetting
\usepackage{booktabs}       % professional-quality tables
\usepackage{amsfonts}       % blackboard math symbols
\usepackage{nicefrac}       % compact symbols for 1/2, etc.
\usepackage{microtype}      % microtypography
\usepackage{xcolor}         % colors
\usepackage{booktabs}
\usepackage{multirow}
\usepackage{graphicx}
\usepackage{colortbl}
\usepackage{amsmath}
\usepackage{pifont}
\usepackage[dvipsnames]{xcolor} % Required for \textcolor
\usepackage{amssymb}      % Optional: Alternative checkmarks
\newcommand{\cmark}{\ding{51}}
\newcommand{\xmark}{\ding{55}}
\usepackage{tabularx}
\usepackage{longtable}
\usepackage{booktabs}
\usepackage{supertabular}
\usepackage{array}
\usepackage{subcaption}
\usepackage{threeparttable}

\def\BibTeX{{\rm B\kern-.05em{\sc i\kern-.025em b}\kern-.08em
    T\kern-.1667em\lower.7ex\hbox{E}\kern-.125emX}}
\usepackage{balance}
\begin{document}

\newcommand{\modelname}{\textbf{\textsc{Jastin}}}

\title{\modelname:  Aligning LLMs for Zero-Shot Audio and Speech Evaluation via Natural Language Instructions}

\author{Leying Zhang, Bowen Shi, Haibin Wu, Bach Viet Do, Yanmin Qian \IEEEmembership{Senior Member, IEEE}
\thanks{Leying Zhang and Yanmin Qian are with the Auditory Cognition and Computational Acoustics Lab, School of Computer Science\& MoE Key Laboratory of Artificial Intelligence, AI Institute, Shanghai Jiao Tong University, Shanghai, 200240 P. R. China (e-mail:\{zhangleying, yanminqian\}@sjtu.edu.cn).

Bowen Shi, Haibin Wu and Bach Viet Do are independent researchers, United States (e-mail:bshi@ttic.edu, f07921092@ntu.edu.com, bachdo@meta.com)}}

\markboth{Journal of \LaTeX\ Class Files,~Vol.~18, No.~9, September~2020}%
{How to Use the IEEEtran \LaTeX \ Templates}

\maketitle

\begin{abstract}
The rapid advancement of generative audio models has outpaced the development of robust evaluation methodologies. Existing objective metrics and general multimodal large language models (MLLMs) often struggle with domain generalization, zero-shot capabilities, and instructional flexibility. To address these bottlenecks, we propose \modelname{}, a generalizable, instruction-driven audio evaluation framework that formulates audio assessment as a self-instructed reasoning task. \modelname\ bridges a frozen high-performance audio encoder with a fine-tuned LLM backbone via a trainable audio adapter. To ensure robust zero-shot generalization, we introduce a comprehensive instruction following data preparation pipeline, incorporating Multi-Source, Multi-Task, Multi-Calibration, and Multi-Description data. Experimental results demonstrate that \modelname\ achieves state-of-the-art Pearson and Spearman correlations with human subjective ratings. It consistently outperforms general MLLMs across speech, sound, music, and out-of-domain evaluation tasks without the need for task-specific retraining.
\end{abstract}


\section{Introduction}

\IEEEPARstart{T}{he} rapid advancement of generative model has led to high-fidelity synthesis across various audio domains, including text-to-speech (TTS), music generation, and environmental sound synthesis~\cite{le2024voicebox, ju2024naturalspeech,zhang2024covomix,zhang2025advanced}. However, the development of robust evaluation methodologies has not kept pace with these generative capabilities. Traditionally, human listening studies, such as Mean Opinion Scores (MOS) or MUSHRA tests~\cite{varadhanrethinking,zhangcovomix2}, have served as the gold standard for assessing subjective quality. Yet, these studies are prohibitively expensive, time-consuming, and difficult to scale for iterative model development.

To automate this evaluation process, numerous objective metrics have been proposed.These approaches can be broadly categorized into traditional non-LLM metrics and emerging LLM-as-a-Judge frameworks.

Traditional signal-processing based metrics, such as PESQ, STOI, and SDR~\cite{rix2001perceptual, jensen2016algorithm, Performance-Vincent2006} remain staples for speech and audio assessment. More recently, neural-network-based metrics like NISQA, UTMOS, DNSMOS, and AES~\cite{mittag2021nisqa,baba2024utmosv2,reddy2021dnsmos, tjandra2025meta} have been developed to better simulate human perception of speech/audio quality.

Further advancing this field, LLM-as-a-Judge frameworks~\cite{zheng2023judging} leverage the sophisticated reasoning of foundational models to score audio via text prompts. By either converting audio into descriptive captions or employing multimodal LLMs (MLLMs) such as Gemini 3 Pro~\cite{gemini} or GPT-4o~\cite{gpt4o} for direct processing, these frameworks evaluate nuanced dimensions including semantic alignment, acoustic fidelity, stylistic consistency, and captioning accuracy~\cite{qiu2026audiocapbench, zhang2026deepasmr}. 

Furthermore, specialized frameworks built on pretrained MLLMs, such as AudioJudge, SpeechJudge, QualiSpeech and SpeechEval ~\cite{manakul2025audiojudge, zhang2025speechjudge, wang2025qualispeech,wang2025speechllm}, demonstrate that large audio models can be prompted to assess specialized speech characteristics such as pronunciation, naturalness, and emotional prosody, effectively automating the role of expert annotators.

Despite these advancements, existing objective models, both traditional and LLM-based, currently face three critical bottlenecks:

First, most traditional metrics suffer from narrow domain applicability. For instance, PESQ is unsuitable for music, while Fréchet Audio Distance (FAD)~\cite{gui2024adapting} cannot adequately evaluate speech, leading to a fragmented evaluation pipeline. Even state-of-the-art (SOTA) neural metrics like Audiobox-Aesthetics (AES)~\cite{tjandra2025meta} often fail to generalize to unseen tasks, making it difficult to determine the most appropriate metric for a given scenario. Moreover, these metrics lack the contextual flexibility to account for the subjectivity of human preference, where the same audio may be judged differently based on specific user descriptions or scenarios.

Second, general MLLMs show inconsistent performance. While general-purpose multimodal models like Gemini 3 Pro or GPT-4o show promising zero-shot task generalization, their performance in specialized audio evaluation remains inconsistent and often fails to meet the precision required for rigorous assessment~\cite{shen2026gsrm}.
 
Third, specialized LLM-based judges often struggle with zero-shot generalization and lack instructional flexibility. These models are also task-specialized, and they rely on rigid prompt templates, making them fragile to slight wording changes. Furthermore, they are typically restricted to fixed scoring scales (e.g., 1–5) and cannot dynamically adjust to alternative user requirements (e.g., a 1–100 scale), limiting their utility in diverse, real-world settings.

To address these challenges, we propose \modelname{}, an LLM-as-a-\textbf{J}udge framework for Zero-Shot \textbf{A}udio and \textbf{S}peech Evaluation \textbf{T}asks via \textbf{I}nstructional \textbf{N}atural Language. It is a generalizable, instruction-driven evaluation framework that moves beyond static, domain-specific metrics by treating audio assessment as a self-instructed task. The architecture comprises a frozen high-performance audio encoder, a trainable audio adapter, and a fine-tuned LLM backbone. This framework introduces four key innovations:

\begin{enumerate}
   
\item \textbf{Unified Generalization:} A single, comprehensive model for zero-shot single-turn speech evaluation task, and is capable of evaluating speech, music, and sound effects without the need for task-specific retraining.

\item \textbf{Comprehensive Data Preparation:} We employ a heterogeneous data preparation pipeline: Multi-source, Multi-task (incorporating human-labeled, pseudo-labeled, and proxy-task data across 24 tasks), Multi-calibration, and Multi-description (utilizing templates for calibration extension and LLMs for description paraphrasing).

\item \textbf{Instructional Robustness:} By employing a self-instructed training paradigm, \modelname\ achieves a critical balance between semantic sensitivity and lexical robustness. It flexibly adapts its behavior to distinct changes in evaluation rules and calibration scales, yet maintains highly consistent scoring when prompts are merely rephrased without altering the core intent.

\item \textbf{Human-Centric Alignment:} Experimental results demonstrate that \modelname\ achieves State-of-the-Art (SOTA) correlations with human subjective ratings compared to both specialized objective metrics and general-purpose, closed-source LLMs.
 
\end{enumerate}

The remainder of this paper is organized as follows: Section \ref{sec:related_work} reviews the relevant literature. Section \ref{sec:methodology} describes our proposed methodology, and Section \ref{sec:data} details our data preparation pipeline. We outline the experimental setup in Section \ref{sec:setup} and present a comparative analysis of our models against established baselines in Section \ref{sec:main-result}. Finally, Section \ref{sec:ablation-limitation} provides an ablation study and discusses failure cases and current limitations.

\section{Related Work}
\label{sec:related_work}
\subsection{Traditional non-LLM Metrics}
Prior to the rise of LLMs, automated audio evaluation relied primarily on signal processing techniques and task-specific neural architectures. For speech synthesis and enhancement, reference-based metrics such as PESQ and STOI have long served as the standard for quantifying signal degradation and intelligibility.

To bridge the gap between objective computation and subjective perception, recent research has shifted toward neural-based MOS prediction. Frameworks such as NISQA, DNSMOS, UTMOS, and UrgentMOS \cite{mittag2021nisqa, reddy2021dnsmos, baba2024utmosv2, wang2026urgentmos} leverage deep learning to approximate human auditory judgment. SAM-Audio-Judge~\cite{wang2026samaudiojudgeunified} is specifically designed to evaluate audio separation without human intervention. Expanding beyond speech, AES~\cite{tjandra2025meta} utilizes a WavLM-based architecture to assess audio quality across four fixed perceptual axes: Production Quality, Complexity, Enjoyment, and Usefulness. This approach enables reference-free evaluation across diverse domains, including music and sound effects.

Despite their utility, these objective metrics remain constrained by fixed calibrations and are limited to providing numerical outputs on hard-coded, predetermined axes. Furthermore, they lack the contextual flexibility to open-ended user descriptions or custom, task-specific evaluation criteria.

\subsection{LLM-as-a-Judge Frameworks}
Leveraging the zero-shot capabilities of MLLMs, recent studies have directly employed models such as Gemini~\cite{comanici2025gemini25,gemini}, GPT-4o~\cite{gpt4o}, and Qwen3-Omni~\cite{xu2025qwen3} for automated evaluation. For instance, these foundational models have been utilized to assess the music perception~\cite{carone2026llms}, evaluate the overall quality of synthesized speech~\cite{zhang2026deepasmr}, and to benchmark general audio-language reasoning capabilities~\cite{yang2025towards}. While these models demonstrate baseline potential through custom prompting strategies, they often exhibit a performance gap when compared to symbolic reasoning or specialized objective metrics, highlighting the need for more robust, instruction-driven frameworks.

To address this limitation, researchers have focused on refining and adapting existing MLLMs to enhance evaluation accuracy. AudioJudge~\cite{manakul2025audiojudge} decomposes assessment into specialized judges for lexical and paralinguistic features, while ARECHO~\cite{shi2025arecho} employs autoregressive dependency modeling for multi-metric speech assessment. To improve descriptive granularity, QualiSpeech~\cite{wang2025qualispeech} provides detailed noise and distortion analysis via a quality-focused dataset. Other efforts prioritize reasoning and interpretability: SpeechEval~\cite{wang2025speechllm} and GSRM~\cite{shen2026gsrm} utilize chain-of-thought (CoT) reasoning to provide explainable judgments, while SpeechJudge~\cite{zhang2025speechjudge} and Kosteno et al.~\cite{kostenok2026calibration} employ post-training and reinforcement learning, respectively, to align models with human perception. Additionally, ALLD~\cite{chen2025audio} utilizes LLM distillation to refine information extraction from raw speech.

An alternative paradigm bypasses native audio LLMs by conducting evaluations through pure text LLMs and ASR model. SpeechQualityLLM~\cite{monjur2025speechqualityllm} achieves this by coupling an audio encoder with a text LLM using template-based Q\&A pairs. Similarly, TRACE~\cite{chandra2026hearing} utilizes a two-stage approach to unlock the reasoning power of text-only models, enabling cost-efficient and better human-aligned evaluation than MLLM without requiring a native speech-capable backbone. 

Despite all these advancements, existing models are often confined to specific metrics or templates, limiting their generalization to unseen tasks.

\section{JASTIN Audio Evaluation Framework}
\label{sec:methodology}

\begin{figure*}[htbp]
    \centering 
\includegraphics[width=1.0\textwidth]{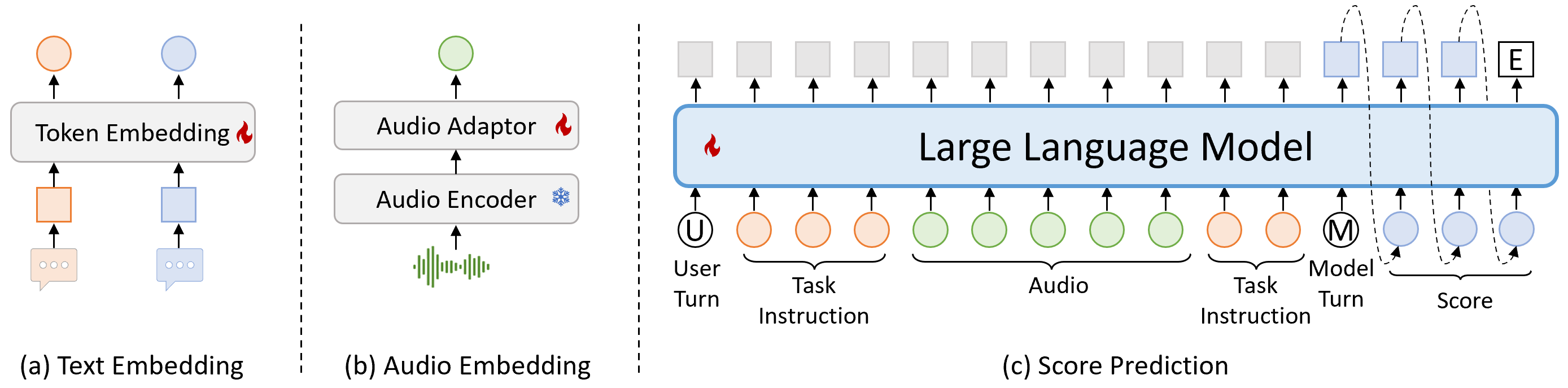} 
    \caption{Pipeline of our proposed framework \modelname} 
    \label{fig:pipeline} 
\end{figure*} 

The primary objective of our framework is to transform audio evaluation from a static, fixed-metric regression problem into an instruction-driven, semantically-sensitive and lexically-robust framework that is generalizable to unseen tasks. We achieve this by bridging high-resolution acoustic representations, extracted via a pre-trained audio encoder, with the advanced linguistic and reasoning capabilities of a LLM.

\subsection{Task Definition}
We formulate audio evaluation as a context-dependent scoring task, mirroring the human annotation process. In real-world scenarios, an evaluator (whether human or artificial) assigns a score based not only on the audio signal itself but also on a specific task description or grading rubric. Consequently, if the evaluation criteria or context shifts, the predicted score must dynamically adapt. Formally, as shown in Eq. \ref{eq:task}, the predicted score $s$ is a function of the evaluation system $f$, the natural language task description $T$, and the input audio $A$:
\begin{equation}
    s = f(T, A)
    \label{eq:task}
\end{equation}

\subsection{Pipeline}
The architecture of \modelname\ is designed to process multi-modal inputs effectively while maintaining computational efficiency. Unlike traditional objective metrics that only ingest raw audio, \modelname\ accepts a multimodal tuple $(T, A)$, where $T$ is the natural language instruction.

To achieve strong instructional robustness, we employ an LLM-driven data augmentation strategy (detailed in Section \ref{sec:data-aug}). For each ground-truth score in our training set (e.g., a MOS of 4.2), a teacher LLM generates a diverse set of augmented task descriptions $T$ to simulate varied user phrasing.

As shown in Figure \ref{fig:pipeline}(a), the task instruction $T$ is tokenized and embedded via the LLM's vocabulary space. The raw audio $A$ is mapped directly to continuous audio embeddings $Z = \phi(E(A))$, bypassing discrete acoustic tokenization. Here, $E$ denotes a frozen, pre-trained audio encoder that extracts robust acoustic features, and $\phi$ represents a lightweight adapter network. This adapter functions as a projection layer, bridging the modality gap between the continuous audio embedding space and the discrete text token space of the LLM.

 As illustrated in Figure \ref{fig:pipeline}(c), we adopt a chat-template input format that allows audio and task instructions to be interleaved. User and model turns are initialized by specific start tokens.  The user-turn context is formatted as $X_{\text{user}} = [\tau_{\text{user}}, T_1, Z, T_2]$, where $\tau_{\text{user}}$ is a specialized user-turn token, and $T_1, T_2$ denote the embedded segments of the task instruction~\footnote{Example: You are a helpful evaluator. Your task is to evaluate the content enjoyment score of an audio waveform on a scale from 1 to 10. This score focuses on the subject quality of an audio piece. It is a more open-ended axis, some aspects might include emotional impact, artistic skill, artistic expression, as well as subjective experience, etc. The higher the score, the more enjoyable the audio is. <audio>. Now, please predict the score of this waveform. }. This structure facilitates flexible multi-modal interleaving, concluded by a specific prompt (e.g., 'Now, please predict the score') to elicit the final prediction.

Let $Y$ represent the target score, formatted as a sequence of text embeddings. The model processes this as the response turn $X_{\text{score}} = [\tau_{\text{model}}, Y]$, where $\tau_{\text{model}}$ denotes the model-turn token. The final unified sequence ingested by the network is the concatenation $X = [X_{\text{user}}, X_{\text{score}}]$.

The model is trained to predict the target score sequence $Y$ autoregressively. Given the interleaved multimodal context $X_{\text{user}}$, we minimize the negative log-likelihood of the target tokens as in Eq. \ref{eq:loss}:

\begin{equation}
  \mathcal{L} = -\sum_{t=1}^{N} \log P(y_t \mid X_{\text{context}}, y_{<t}; \theta) 
   \label{eq:loss}
\end{equation}
where $y_t$ represents the $t$-th token of $Y$, $N$ is the sequence length of the target score, and $\theta$ encapsulates the trainable parameters of both the adapter $\phi$ and the LLM. Notably, the loss is computed exclusively over the target response tokens $Y$; the user turn $X_{\text{user}}$ is masked during loss calculation. By formulating the input as a unified conversational sequence, the network naturally learns to attend to the acoustic features exactly where they are referenced, subsequently reasoning through the textual rubric to generate the final numerical score.

Finally, our framework is fundamentally model-agnostic. While we instantiate it with specific architectures in our experiments, both the audio encoder and the LLM backbone can be seamlessly substituted with better alternative foundation models in the future.

\begin{figure*}[htbp]
    \centering 
\includegraphics[width=1.0\textwidth]{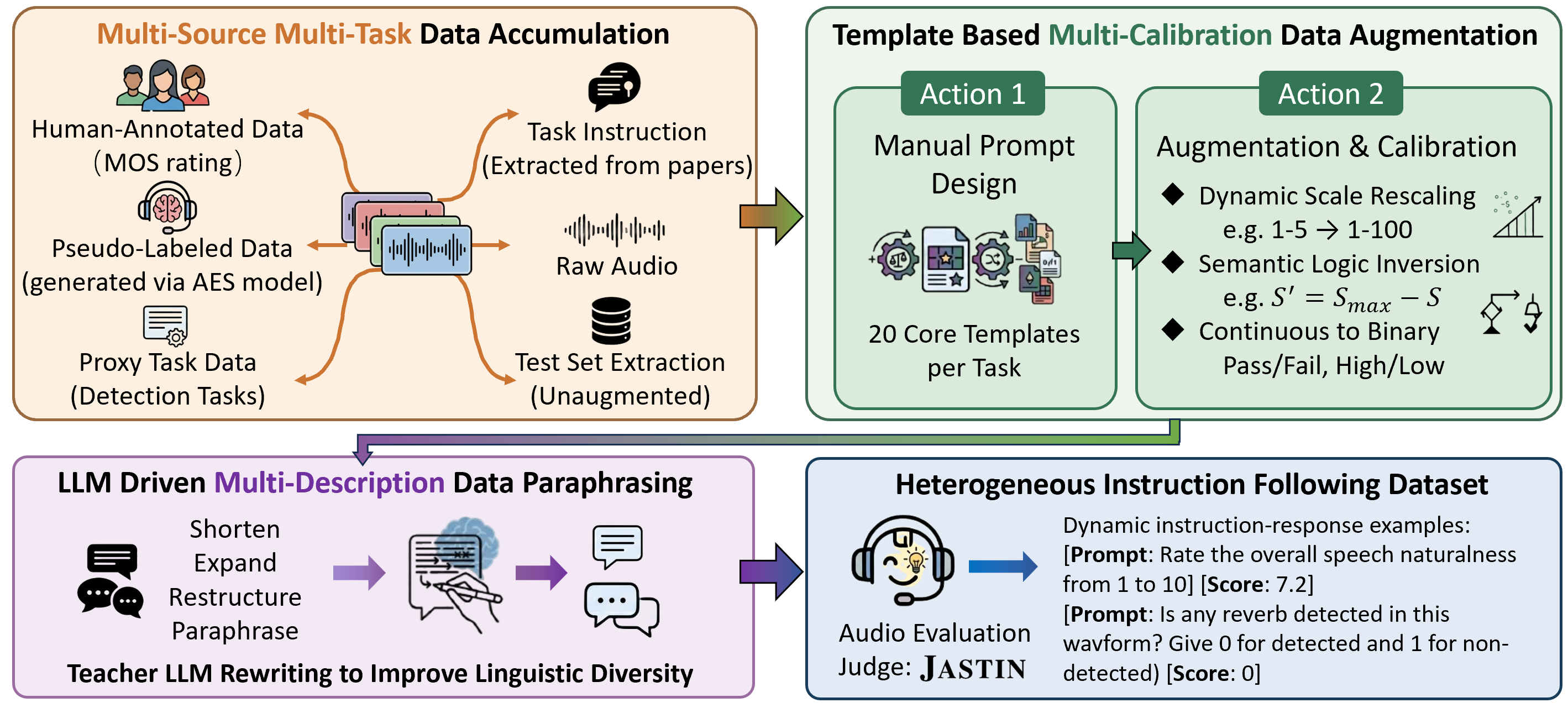} 
    \caption{Data preparation pipeline of our proposed framework \modelname} 
    \label{fig:data-pipeline} 
\end{figure*}

\subsection{Model Architecture}

The audio encoder serves as the critical bridge between raw acoustic signals and the text-based LLM backbone. It needs to generalize broadly across speech, music, and general audio.  Therefore, we utilize the PE-A-Frame-base model\footnote{https://huggingface.co/facebook/pe-a-frame-base}\cite{vyas2025pushing} as our primary audio encoder and generate an audio embedding of dimension 1024. The PE-A-Frame-base is a specialized variant of Perception Encoder, optimized for high-resolution audio understanding. Built upon a multimodal architecture, it leverages frame-level contrastive learning to align temporal audio segments with linguistic descriptions. Unlike global encoders that summarize entire clips, PE-A-Frame focuses on fine-grained temporal dynamics, enabling precise localization of acoustic events. 

To align modalities, an audio adapter consisting of a linear projection followed by a bottleneck residual adapter is employed. This adapter features a 4:1 compression ratio and GELU activations, mapping audio features into a normalized text embedding space of dimension 1024. 

For the language backbone, we adopt Llama-3.2-3B\footnote{https://huggingface.co/meta-llama/Llama-3.2-3B-Instruct}~\cite{llama322024}, which serves as the core reasoning engine. The projected audio features are treated as a sequence of continuous embeddings and interleaved within the chat template. By keeping the audio encoder weights frozen and training the adapter and LLM, we align the high-resolution temporal features of the audio with the semantic space of the language model, allowing for sophisticated, instruction-driven audio evaluation.

\section{Data Preparation for JASTIN Optimization}
\label{sec:data}
To train a robust and generalizable audio evaluation judge, we curate a heterogeneous dataset characterized by multi-source, multi-task, multi-calibration, and multi-description attributes. The data preparation pipeline is shown in Figure \ref{fig:data-pipeline} with detailed introduction in the following subsections.

\subsection{Multi-Source Data Collection}
 \label{sec:data-source}
To ensure our model achieves broad generalization across diverse acoustic scenarios, we aggregate English speech, sound, and music data from three distinct sources:

\textbf{1. Human-Annotated Ground Truth:} This consists of datasets where human evaluators have explicitly provided scalar ratings for audio quality, forming the backbone of the model’s understanding of subjectivity and naturalness. We utilize multiple datasets with 24,000 utterances, containing both synthetic and real audio, including BVCC~\cite{cooper21_ssw}, QualiSpeech~\cite{wang2025qualispeech}, SpeechEval~\cite{wang2025speechllm}, and UrgentMOS~\cite{wang2026urgentmosunifiedmultimetricpreference}.

\textbf{2. Pseudo-Labeled Data for Scale Extension: } To expand our training corpus, we collected over 80,000 utterances from public datasets (LibriTTS~\cite{zen2019libritts}, Expresso~\cite{nguyen2023expresso}, CommonVoice~\cite{ardila2020common}, EARS~\cite{richter2024ears}, AudioSet~\cite{gemmeke2017audio}, FreeSound~\cite{fonseca2017freesound}, MusicCaps~\cite{agostinelli2023musiclm}, MUSDB18~\cite{rafii2017musdb18}). We then utilized the public AES model~\cite{tjandra2025meta} to generate pseudo-labels across dimensions such as CE, CU, PC, and PQ for each utterance.

\textbf{3. Proxy Data for Broad Generalization: } To prevent the model from overfitting to a narrow definition of ``quality," we incorporate detection proxy tasks~\footnote{Example: Does the voice exhibit a tone of amusement? Please respond with 1 if amusement is present, 0 otherwise. <audio> Now, please predict the score of this waveform.}. These tasks teach the model to map specific acoustic characteristics to scalar confidence scores or classifications, which the LLM interprets as evaluation dimensions. We define these proxy tasks across several domains, including child speech, emotion and style detection, reverberation detection and music distortion detection. 

We utilize the ground-truth labels of ChildSpeech~\footnote{https://huggingface.co/datasets/TomRoma/Child\_Speech\_dataset\_Whisper}, Expresso~\cite{nguyen2023expresso} and CHAINs~\cite{cummins2006chains} for emotion, style and child-speech detection. Moreover, we synthesize distorted data using samples from LibriSpeech~\cite{librispeech}, MusicCaps~\cite{agostinelli2023musiclm} and NCSSD~\cite{NCSSD}. Specifically, we achieve reverberation detection by manually applying reverberation effects. For music distortion detection, we introduce artifacts such as anomalous sounds, human noise, and sudden silence. We collected a total of 43,500 utterances as our proxy-task data.

All data and tasks are converted into a universal format consisting of task instruction, audio, additional instructions, and score, as shown in Figure~\ref{fig:data-pipeline}. They are unified under a single training objective and inference procedure.

\subsection{Multi-Task Data Preparation}
\label{sec:data-task}
For each data source, we evaluate multiple tasks. By including diverse objectives, the model learns to act as a multi-faceted judge, dynamically adapting its evaluation criteria based on the input prompt. In total, we incorporate 24 tasks encompassing various metrics for synthesized and converted speech, such as naturalness, prosody, emotion, and distortion. A comprehensive task list is provided in supplementary materials~\footnote{https://github.com/vivian556123/Jastin/blob/main/prompts-and-tasks.html}.

Consolidating multi-task data from disparate sources presents a critical challenge: the inherent inconsistency in task definitions across datasets. To address this, we condition the model on detailed task descriptions rather than generic task names. Consequently, scores for similar tasks (e.g., ``distortion" or ``overall quality") are explicitly grounded in their specific rubric definitions rather than assumed generic categories.

For example, QualiSpeech uses a flat structure targeting low-level acoustic features, strictly separating noise (environmental interference) and distortion (the voice itself) into distinct 1–5 scoring metrics. Conversely, SpeechEval employs a hierarchical structure geared toward human perception, grouping all Noise and machine artifacts into a single distortion score where 1 indicates severe artifacts and 5 indicates clean audio.

\subsection{Multi Calibration Multi Description LLM-driven Data Augmentation}
\label{sec:data-aug}
Our data preparation follows a structured three-step workflow to transform raw audio data into instruction-following evaluation pairs.

\subsubsection{\textbf{Step 1: Data Accumulation and Proxy Task Synthesis}}

As introduced in Section \ref{sec:data-source} and Section \ref{sec:data-task}, we aggregate a heterogeneous collection of audio datasets covering speech, music, and general sound. We extract the official task descriptions from the original source papers to serve as our default prompts. Beyond standard quality metrics (e.g., MOS), we design Proxy Tasks to teach the model semantic and acoustic reasoning. Note that the entirety of our test set is derived exclusively from this unaugmented pool.

\subsubsection{\textbf{Step 2: Template-Based Multi-Calibration Augmentation}}

For each task identified in Step 1, we manually design more than 20 core prompt templates. To ensure the model does not overfit to a specific numerical range or direction, we apply the following augmentations:

First, we dynamically rescale target scores. For instance, a ground-truth score of 4.2 on a 1–5 scale is mapped to 8.4 on a 1–10 scale, or 84 on a 1–100 scale.
Second, we invert the semantic logic of the prompt (e.g., changing ``Rate the amount of noise" to ``Rate the clarity of the signal") and adjust the target score accordingly ($S' = S_{max} - S$)
Third, we convert continuous scores into binary ``pass/fail" or ``high/low" classifications, helping the model learn discrete decision boundaries.

\subsubsection{\textbf{Step 3: LLM-Driven Multi-Description Paraphrasing}}

To ensure the model is robust to natural language variations, we employ a teacher LLM to rewrite the task descriptions generated in Step 2. This process introduces linguistic diversity, transforming formal requests into casual questions or highly technical rubrics. This guarantees that the model learns the underlying intent of the evaluation rather than memorizing specific keyword triggers. We utilize various prompting strategies to shorten, expand, restructure, and heavily paraphrase the original text. 

Importantly, Steps 2 and 3 are applied strictly to the training data.  The test set remains the same as in other papers and baselines to ensure rigorous and fair evaluation.   We open-source the model design, inference scripts, data-processing scripts, and all the templates, task descriptions, and prompts to promote further research\footnote{https://github.com/vivian556123/Jastin}.

\section{Experimental Setup}
\label{sec:setup}
\subsection{Training Configuration}

Training is conducted on eight NVIDIA A100 GPUs for 6,000 steps (about 24 hours). We use a per-GPU batch size of 6 and 8 gradient accumulation sub-steps, resulting in an effective total batch size of 384 samples per step. We normalize the data to keep 2 digits of the final predicted score. We apply gradient clip of 0.2. We utilize Polynomial Decay Scheduler with lr=1e-5 and warmup 1000 steps with AdamW optimizer. We employ early stopping by monitoring the Pearson Correlation Coefficient on the AES Speech PQ metric within the validation set. For inference, we do not do sampling and the max generation length is 100. 

\subsection{Test Sets and Evaluation Metrics}
We evaluate our model across five various datasets, assessing different aspects of audio and speech quality:

\subsubsection{QualiSpeech~\cite{wang2025qualispeech}}: Evaluated using six metrics: noise, Distortion (Dist.), Continuity (Cont.), Listening Effort (Listen.), Naturalness (Nat.), and Overall Quality (Ovrl.).

\subsubsection{SpeechEval~\cite{wang2025speechllm}}: Evaluated using seven metrics: Overall Quality (Ovrl.), Intelligibility (Int.), Distortion (Dist.), Dynamic Range (Dyn.), Emotional Impact (Emo.), Artistic Expression (Art.), and Subjective Experience (Subj.).

\subsubsection{AES~\cite{tjandra2025meta}}: Evaluated using four metrics: Content Enjoyment (CE), Content Usefulness (CU), Production Complexity (PC), and Production Quality (PQ).

\subsubsection{AudioMOS2025~\cite{huang2025audiomos}}: Evaluated using three metrics: Overall Musical Quality (M-Ovrl.), Music-Textual Alignment (M-TA.), and MOS prediction with synthesized speech at different sampling rates (SynMOS). We treat this as an out-of-domain test set with unseen data and task descriptions. 

\subsubsection{DeepASMR~\cite{zhang2026deepasmr}}: Evaluated via overall quality prediction (AsmrMOS). Similar to AudioMos2025, this serves as an out-of-domain test set with unseen data and task descriptions.

\begin{table*}[htbp]
\centering
\begin{threeparttable}
\caption{Comparison between our \modelname\ and baseline models on Speech-only Datasets.}
\label{tab:qualispeech_and_speecheval_main}
\begin{tabular}{l| cccccc |ccccccc}
\toprule
\multirow{2}{*}{\textbf{Model}} & \multicolumn{6}{c|}{\textbf{QualiSpeech}} & \multicolumn{7}{c}{\textbf{SpeechEval}} \\
\cmidrule(lr){2-7} \cmidrule(lr){8-14}
 & Noise & Dist. & Cont. & Listen. & Nat. & Ovrl. & Ovrl. & Int. & Dist. & Dyn. & Emo. & Art. & Subj. \\
\midrule
\rowcolor[gray]{.9} \multicolumn{14}{c}{\textbf{Pearson Correlation (PCC $\uparrow$)}} \\
\midrule
QualiSpeech$^*$~\cite{wang2025qualispeech}  & \textbf{0.686} & 0.518 & 0.459 & 0.475 & 0.486 & \textbf{0.572} & - & - & - & - & - & - & - \\
SpeechEval$^*$~\cite{wang2025speechllm}   & - & - & - & - & - & & 0.520 & 0.505 & 0.592 & 0.329 & 0.434 & 0.378 & 0.456 \\
\midrule
AES-CE~\cite{tjandra2025meta}       & 0.223 & 0.497 & 0.400 & \textbf{0.505} & 0.459 & 0.513 & 0.661 & 0.616 & \underline{0.705} & \underline{0.512} & \underline{0.563} & \underline{0.502} & \underline{0.564} \\
AES-CU~\cite{tjandra2025meta}         & 0.193 & 0.402 & 0.370 & 0.425 & 0.346 & 0.411 & 0.573 & 0.533 & 0.625 & 0.502 & 0.522 & 0.456 & 0.462 \\
AES-PC~\cite{tjandra2025meta}         & -0.575 & -0.017 & -0.100 & -0.203 & -0.034 & -0.174 & -0.141 & -0.138 & -0.182 & -0.163 & -0.145 & -0.144 & -0.079 \\
AES-PQ~\cite{tjandra2025meta}         & 0.182 & 0.404 & 0.328 & 0.398 & 0.350 & 0.405 & 0.602 & 0.563 & 0.677 & \textbf{0.525} & 0.553 & 0.489 & 0.483 \\
UTMOS~\cite{baba2024utmosv2}        & 0.174 & 0.482 & 0.271 & 0.444 & 0.448 & 0.482 & \textbf{0.748} & \textbf{0.716} & \textbf{0.740} & 0.524 & \textbf{0.569} & \textbf{0.519} & \textbf{0.623} \\
NISQA~\cite{mittag2021nisqa}        & 0.315 & 0.290 & 0.239 & 0.335 & 0.266 & 0.336 & 0.620 & 0.584 & 0.611 & 0.419 & 0.503 & 0.467 & 0.515 \\
\midrule
Gemini-3-Pro$^+$~\cite{gemini} & 0.381 & \underline{0.560} & \textbf{0.483} & 0.475 & \underline{0.530} & 0.520 & 0.497 & 0.463 & 0.529 & 0.176 & 0.306 & 0.289 & 0.449 \\
Gemini-2.5-Pro$^+$~\cite{comanici2025gemini25} & 0.406 & 0.424 & 0.434 & 0.430 & 0.406 & 0.383 & 0.343 & 0.295 & 0.226 & 0.055 & 0.197 & 0.168 & 0.255 \\
Gemini-2.5-Flash$^+$~\cite{comanici2025gemini25} & 0.240 & 0.401 & 0.305 & 0.392 & 0.308 & 0.392 & 0.437 & 0.229 & 0.415 & 0.303 & 0.166 & 0.209 & 0.353 \\
Qwen3-Omni~\cite{xu2025qwen3}  & 0.277 & 0.263 & 0.362 & 0.347 & 0.367 & 0.384 & 0.407 & 0.406 & 0.241 & 0.081 & 0.063 & 0.125 & 0.169 \\
Qwen2-Audio~\cite{chu2024qwen2} & -0.081 & -0.047 & -0.068 & 0.005 & 0.018 & 0.042 & 0.097 & 0.064 & 0.018 & -0.003 & -0.003 & 0.113 & 0.056 \\
AudioFlamingo3~\cite{goel2025audioflamingo} & 0.019 & -0.190 & 0.034 & -0.146 & 0.158 & 0.113 & 0.000 & 0.033 & 0.093 & 0.156 & -0.055 & -0.038 & 0.112 \\
\midrule
\modelname\         & \underline{0.668} & \textbf{0.561} & \underline{0.477} & \underline{0.497} & \textbf{0.604} & \underline{0.549} & \underline{0.662} & \underline{0.655} & 0.690 & 0.481 & 0.564 & 0.509 & 0.534 \\
\midrule
\rowcolor[gray]{.9} 
\multicolumn{14}{c}{\textbf{Spearman Correlation (SRCC $\uparrow$)}} \\
\midrule
AES-CE~\cite{tjandra2025meta}       & 0.192 & 0.496 & 0.387 & \textbf{0.489} & 0.457 & 0.515 & \underline{0.657} & \underline{0.616} & \textbf{0.713} & 0.508 & \textbf{0.577} & \underline{0.505} & \underline{0.550} \\
AES-CU~\cite{tjandra2025meta}       & 0.175 & 0.384 & 0.349 & 0.404 & 0.337 & 0.405 & 0.539 & 0.523 & 0.626 & \underline{0.522} & 0.506 & 0.434 & 0.421 \\
AES-PC~\cite{tjandra2025meta}       & -0.421 & -0.021 & -0.221 & -0.205 & -0.019 & -0.143 & -0.042 & -0.069 & -0.153 & -0.143 & -0.137 & -0.122 & -0.011 \\
AES-PQ~\cite{tjandra2025meta}       & 0.158 & 0.396 & 0.322 & 0.391 & 0.346 & 0.405 & 0.592 & 0.570 & 0.678 & \textbf{0.535} & 0.548 & 0.483 & 0.467 \\
UTMOS~\cite{baba2024utmosv2}        & 0.155 & 0.495 & 0.281 & 0.462 & 0.458 & 0.500 & \textbf{0.745} & \textbf{0.700} & \textbf{0.717} & 0.506 & 0.573 & \textbf{0.520} & \textbf{0.624} \\
NISQA~\cite{mittag2021nisqa}        & 0.261 & 0.284 & 0.244 & 0.312 & 0.262 & 0.329 & 0.630 & 0.573 & 0.605 & 0.420 & 0.515 & 0.470 & 0.524 \\
\midrule
Gemini-3-Pro$^+$~\cite{gemini} & \underline{0.306} & \textbf{0.570} & \textbf{0.458} & \underline{0.472} & \underline{0.538} & \textbf{0.568} & 0.496 & 0.477 & 0.524 & 0.181 & 0.315 & 0.281 & 0.456 \\
Gemini-2.5-Pro$^+$~\cite{comanici2025gemini25} & 0.304 & 0.407 & 0.396 & 0.332 & 0.390 & 0.335 & 0.302 & 0.236 & 0.220 & -0.027 & 0.186 & 0.140 & 0.250 \\
Gemini-2.5-Flash$^+$~\cite{comanici2025gemini25} & 0.232 & 0.399 & 0.265 & 0.366 & 0.305 & 0.386 & 0.440 & 0.217 & 0.426 & 0.279 & 0.182 & 0.198 & 0.339 \\
Qwen3-Omni~\cite{xu2025qwen3}  & 0.230 & 0.246 & 0.369 & 0.303 & 0.368 & 0.384 & 0.414 & 0.391 & 0.241 & 0.127 & 0.055 & 0.120 & 0.198 \\
Qwen2-Audio~\cite{chu2024qwen2} & -0.029 & -0.036 & -0.126 & -0.038 & 0.024 & 0.038 & 0.117 & -0.029 & 0.005 & 0.022 & 0.069 & 0.126 & 0.053 \\
AudioFlamingo3~\cite{goel2025audioflamingo} & 0.015 & -0.184 & 0.049 & -0.150 & 0.161 & 0.119 & 0.012 & 0.040 & 0.099 & 0.149 & -0.061 & -0.019 & 0.144 \\
\midrule
\modelname\          & \textbf{0.630} & \textbf{0.570} & \underline{0.398} & 0.466 & \textbf{0.624} & \underline{0.555} & \textbf{0.670} & \textbf{0.638} & \underline{0.685} & 0.429 & \underline{0.567} & 0.506 & 0.542 \\
\bottomrule
\end{tabular}
\begin{tablenotes}
    \footnotesize
    \item $^*$ denotes results reported from the official papers. $^+$ denotes models evaluated via API. All other models are inferred with the original code and weights.
    \item The best results are highlighted in \textbf{bold}, while the second-best results are \underline{underlined}.
\end{tablenotes}
\end{threeparttable}
\end{table*}

\subsection{Baselines and Evaluation Metrics}
We compare our proposed \modelname\ framework against three categories of baselines: 
\subsubsection{Non-LLM metrics} We utilize AES model~\cite{tjandra2025meta} with its CE, CU, PC, and PQ metrics, UTMOS~\cite{baba2024utmosv2}, and NISQA~\cite{mittag2021nisqa} as our baselines. 
\subsubsection{General-purpose LLMs} We choose several MLLM models as baselines, including Gemini series ( Gemini-3-Pro, Gemini-2.5-Pro, and Gemini-2.5-Flash)~\cite{gemini}, Qwen series (Qwen3-omni~\cite{xu2025qwen3}, Qwen2-audio~\cite{chu2024qwen2}), and Nvidia's Audio Flamingo3~\cite{goel2025audioflamingo}. 
\subsubsection{Specialized LLMs } We utilize the MLLM fine-tuned on the corresponding QualiSpeech~\cite{wang2025qualispeech} and SpeechEval~\cite{wang2025speechllm} datasets as the specialized LLM baselines, with results reported in the original paper.

To evaluate the accuracy of the predicted scores against human judgments, we report both the Pearson Correlation Coefficient (PCC) and the Spearman Rank Correlation Coefficient (SRCC).  The PCC measures the degree of linear relationship between the predicted and human-rated scores, indicating how well the predictions follow a straight-line trend with the ground truth. The SRCC assesses the monotonic relationship between the two sets of scores by evaluating how consistently the predictions preserve the relative ranking of the samples, regardless of whether the relationship is linear. Higher values for all these coefficients indicate better performance.

We report the PCC results for QualiSpeech~\cite{wang2025qualispeech} and SpeechEval~\cite{wang2025speechllm} as documented in their respective papers. For the Gemini-series models, we utilize the official API. All other baseline models were evaluated by running inference using their original source code and weights. The best results are highlighted in \textbf{bold}, while the second-best results are \underline{underlined}.

\begin{table*}[htbp]
\centering
\begin{threeparttable}
\caption{Comparison between our \modelname\ and baseline models  on AES Speech, Sound and Music dataset}
\label{tab:aes_main}
%\setlength{\tabcolsep}{5pt} 
% \footnotesize 
\begin{tabular}{l | cccc| cccc |cccc}
\toprule
\multirow{2}{*}{\textbf{Model}} & \multicolumn{4}{c|}{\textbf{Speech}} & \multicolumn{4}{c|}{\textbf{Sound}} & \multicolumn{4}{c}{\textbf{Music}} \\
\cmidrule(lr){2-5} \cmidrule(lr){6-9} \cmidrule(lr){10-13}
 & CE & CU & PC & PQ & CE & CU & PC & PQ & CE & CU & PC & PQ \\
\midrule
\rowcolor[gray]{.9} 
\multicolumn{13}{c}{\textbf{Pearson Correlation (PCC $\uparrow$)}} \\
\midrule
AES~\cite{tjandra2025meta}              & \textbf{0.564} & \textbf{0.614} & \underline{0.590} & \textbf{0.730} & \textbf{0.560} & \underline{0.616} & \textbf{0.540} & \textbf{0.609} & \underline{0.748} & \underline{0.690} & \textbf{0.641} & \textbf{0.688} \\ 
UTMOS~\cite{baba2024utmosv2} & 0.286 & 0.306 & -0.148 & 0.360 & -0.054 & -0.108 & 0.011 & -0.108 & -0.141 & -0.144 & 0.082 & -0.108 \\ 
NISQA~\cite{mittag2021nisqa} & 0.311 & 0.366 & 0.223 & 0.433 & 0.043 & -0.013 & -0.153 & -0.009 & -0.078 & -0.020 & -0.201 & 0.003 \\
\midrule
Gemini-3-Pro$^+$~\cite{gemini}   & 0.307 & 0.216 & 0.429 & 0.381 & 0.232 & 0.365 & 0.307 & 0.384 & 0.543 & 0.405 & 0.334 & 0.497 \\
Gemini-2.5-Pro$^+$~\cite{comanici2025gemini25}   & 0.251 & 0.310 & 0.371 & 0.413 & 0.249 & 0.277 & 0.279 & 0.408 & 0.540 & 0.401 & 0.296 & 0.518 \\
Gemini-2.5-Flash$^+$~\cite{comanici2025gemini25} & 0.215 & 0.203 & 0.304 & 0.328 & 0.204 & 0.249 & 0.228 & 0.409 & 0.360 & 0.185 & 0.030 & 0.320 \\
Qwen3-Omni~\cite{xu2025qwen3}      & 0.190 & 0.248 & 0.183 & 0.349 & 0.240 & 0.280 & 0.291 & 0.322 & 0.519 & 0.498 & 0.295 & 0.528 \\
Qwen2-Audio~\cite{chu2024qwen2}    & 0.032 & -0.020 & -0.014 & -0.003 & -0.074 & 0.046 & -0.010 & -0.015 & 0.188 & 0.127 & 0.038 & 0.169 \\
AudioFlamingo3~\cite{goel2025audioflamingo}  & 0.037 & 0.018 & 0.018 & 0.087 & 0.049 & 0.226 & -0.063 & 0.169 & -0.004 & 0.430 & 0.119 & 0.024 \\ \midrule
\modelname\              & \underline{0.531} & \underline{0.594} & \textbf{0.601} & \underline{0.707} & \underline{0.542} & \textbf{0.618} & \underline{0.495} & \underline{0.569} & \textbf{0.749} & \textbf{0.693} & \underline{0.616} & \underline{0.669} \\
\midrule
\rowcolor[gray]{.9} 
\multicolumn{13}{c}{\textbf{Spearman Correlation (SRCC $\uparrow$)}} \\
\midrule
AES~\cite{tjandra2025meta}                  & \textbf{0.536} & \textbf{0.574} & \underline{0.401} & \underline{0.665} & \underline{0.529} & \textbf{0.583} & \textbf{0.561} & \textbf{0.560} & \textbf{0.660} & \textbf{0.645} & \textbf{0.571} & \textbf{0.614} \\ 
UTMOS~\cite{baba2024utmosv2}  & 0.260 & 0.273 & -0.148 & 0.306 & -0.055 & -0.086 & 0.004 & -0.072 & -0.129 & -0.134 & 0.072 & -0.106  \\ 
NISQA~\cite{mittag2021nisqa}  & 0.290 & 0.351 & 0.186 & 0.412 & 0.010 & 0.031 & -0.149 & -0.032 & -0.076 & -0.017 & -0.196 & 0.017 \\
\midrule
Gemini-3-Pro$^+$~\cite{gemini}  & 0.279 & 0.206 & 0.292 & 0.386 & 0.200 & 0.321 & 0.408 & 0.340 & 0.443 & 0.416 & 0.320 & 0.432 \\
Gemini-2.5-Pro$^+$~\cite{comanici2025gemini25}  & 0.243 & 0.265 & 0.285 & 0.421 & 0.243 & 0.214 & 0.323 & 0.381 & 0.409 & 0.373 & 0.292 & 0.399 \\
Gemini-2.5-Flash$^+$~\cite{comanici2025gemini25} & 0.192 & 0.217 & 0.256 & 0.304 & 0.179 & 0.201 & 0.211 & 0.324 & 0.302 & 0.196 & 0.311 & 0.247 \\
Qwen3-Omni~\cite{xu2025qwen3}       & 0.168 & 0.219 & 0.174 & 0.307 & 0.224 & 0.318 & 0.283 & 0.350 & 0.419 & 0.441 & 0.171 & 0.424 \\
Qwen2-Audio~\cite{chu2024qwen2}     & 0.057 & 0.036 & -0.006 & 0.015 & -0.100 & 0.025 & -0.003 & -0.046 & 0.195 & 0.122 & 0.097 & 0.162 \\
AudioFlamingo3~\cite{goel2025audioflamingo} & 0.029 & 0.038 & 0.056 & 0.096 & 0.067 & 0.323 & -0.092 & 0.098 & 0.021 & 0.247 & 0.105 & 0.024 \\ \midrule
\modelname\              & \underline{0.497} & \underline{0.561} & \textbf{0.441} & \textbf{0.666} & \textbf{0.549} & \underline{0.580} & \underline{0.529} & \underline{0.514} & \underline{0.653} & \underline{0.635} & \underline{0.554} & \underline{0.588} \\
\bottomrule
\end{tabular}
\begin{tablenotes}
    \footnotesize
    \item  $^+$ denotes models evaluated via API. All other models are inferred with the original code and weights.
    \item The best results are highlighted in \textbf{bold}, while the second-best results are \underline{underlined}.
\end{tablenotes}
\end{threeparttable}
\end{table*}

\section{Main Results and Analysis}
\label{sec:main-result}

\subsection{Evaluation on Speech-only Dataset}

Table \ref{tab:qualispeech_and_speecheval_main} compares the performance of various models across the QualiSpeech and SpeechEval datasets. We evaluate our proposed \modelname\ against six traditional non-LLM metrics, six general-purpose MLLMs, and two specialized LLM-based evaluators. For QualiSpeech and SpeechEval models, we report PCC results directly from their original paper. For the other baselines, the PCC and SRCC results were obtained through local inference or via API.

We first observe that non-LLM metrics exhibit inconsistent generalization: while they perform adequately on the SpeechEval dataset, their efficacy degrades significantly on QualiSpeech, even for the distortion and overall quality tasks. This discrepancy arises because, despite nominally evaluating the same task, the two datasets focus on entirely different acoustic dimensions. QualiSpeech~\cite{wang2025qualispeech} is explicitly designed around low-level speech perception, whereas SpeechEval strongly prioritizes high-level subjective dimensions~\cite{wang2025speechllm}. Furthermore, conventional non-LLM metrics are typically trained on tasks where their global feature representations naturally align with SpeechEval’s hierarchical ontology.

Among the general MLLMs, the Gemini series and Qwen3-Omni demonstrate the strongest relative performance. The advancement of general MLLM indeed improves the evaluation performance. However, their Pearson correlations coefficient for most metrics remain below 0.50, and their massive parameter counts make them computationally expensive for routine evaluation tasks. 

Specialized LLMs like QualiSpeech and SpeechEval, though not designed for broad generalization, outperform general LLMs on their respective target datasets. This suggests that domain-specific post-training significantly enhances LLM performance for speech evaluation. 

Ultimately, our proposed \modelname\ achieves consistently superior performance, outperforming non-LLM, general LLM, and specialized LLM baselines across almost all metrics in both Pearson and Spearman correlations. These results demonstrate its robustness and highlight its promise as a generalized speech evaluation metric.

\subsection{Evaluation on Sound and Music Dataset}
To evaluate performance across diverse audio domains, including music and general sound, we extend our assessment to the Audiobox Aesthetics (AES) dataset (Table \ref{tab:aes_main}). Our proposed model achieves results comparable to established non-LLM AES baselines, demonstrating that LLM-based architectures can effectively match the performance of traditional, domain-specific systems. In contrast, speech-centric metrics such as UTMOS and NISQA prove unsuitable for non-speech tasks. Their scores do not map intuitively to the four AES metrics, resulting in poor correlation. While Gemini-3-Pro and Qwen3-Omni lead among general-purpose LLMs, they still fall short of the non-LLM baselines and our proposed approach, highlighting a remaining gap in general-purpose audio evaluation.

\subsection{Zero-Shot Generalization on Out-of-Domain Tasks}

A significant advantage of LLM-based evaluators is their potential to generalize to unseen tasks and domains through natural language instructions. To evaluate this, we apply four distinct out-of-domain tasks across diverse audio scenarios: (1) M-TA, assessing text-to-music alignment. (2) M-Ovrl, measuring overall music quality. (3) SynMOS, evaluating Mean Opinion Scores (MOS) for synthesized speech across varying sampling rates. (4) AsmrMOS, focusing on specialized, high-fidelity speech styles such as ASMR. 

The M-TA and AsmrMOS tasks are entirely out-of-domain, differing significantly in both task descriptions and input waveforms. While M-Ovrl and SynMOS assess the overall quality of music and speech across varying sample rates, their unique and unseen task descriptions distinguish them from existing benchmarks like AES, SpeechEval, or the QualiSpeech test sets.

As illustrated in Table \ref{tab:ood-performance}, non-LLM neural evaluators exhibit a lack of consistency when applied to unseen tasks. While these models show localized strengths, their performance is fragmented: CE excels in music-textual alignment (although CE does not accept text as input), UTMOS leads in synthesized speech MOS, and CU proves most effective for ASMR and musical impressions. This specialized behavior highlights a critical flaw, which is the limited semantic transparency and a failure to maintain robust performance across diverse out-of-domain tasks. 

Consequently, these metrics cannot be treated as universal tools. Their lack of adaptability forces a ``trial-and-error" approach, where the user must manually identify a specific model for each new task rather than relying on a generalized evaluation standard.

\modelname\ consistently surpasses the general LLMs across all out-of-domain (OOD) benchmarks, significantly outperforming state-of-the-art general LLMs including Gemini-3-Pro and Qwen3-Omni. In the music domain (M-TA and M-Ovrl), our model exhibits a substantial margin over the baselines, particularly in text-alignment tasks where general LLMs often struggle to correlate acoustic features with textual descriptions.

Furthermore, while general LLMs like Gemini-3-Pro show limited predictive power on synthesized speech and specialized ASMR content, \modelname\ maintains robust correlation coefficients (e.g., 0.496 on SynMOS and 0.297 on AsmrMOS). These results demonstrate that \modelname\ possesses superior zero-shot generalization capabilities, effectively extending its evaluative logic to novel task descriptions and out-of-distribution audio content without the need for task-specific fine-tuning.

\begin{table*}[htbp]
\begin{threeparttable}
\centering
\caption{Comparison between our \modelname\ and baseline models  on Out-of-Domain datasets}
\label{tab:ood-performance}
\begin{tabular}{l |cc| cc| cc| cc}
\toprule
\multirow{2}{*}{\textbf{Model}} &  \multicolumn{2}{c|}{\textbf{Music Textual Alignment}} & \multicolumn{2}{c|}{\textbf{Overall Musical Quality}} & 
\multicolumn{2}{c|}{\textbf{Synthesized Speech MOS}} & \multicolumn{2}{c}{\textbf{ASMR Speech MOS}} \\
\cmidrule(lr){2-3} \cmidrule(lr){4-5} \cmidrule(lr){6-7} \cmidrule(lr){8-9}
& PCC $\uparrow$ & SRCC $\uparrow$ &PCC $\uparrow$ & SRCC $\uparrow$ & PCC $\uparrow$ & SRCC $\uparrow$ & PCC $\uparrow$ & SRCC $\uparrow$  \\
\midrule
AES-CE~\cite{tjandra2025meta} & \textbf{0.509} & \textbf{0.506} & \underline{0.639} & 0.628 &0.512 & 0.460 & 0.179 & 0.156 \\ 
AES-CU~\cite{tjandra2025meta} & 0.471 & 0.495 & 0.627 & \textbf{0.676} & 0.394 & 0.410 & \textbf{0.314} & \underline{0.243} \\ 
AES-PC~\cite{tjandra2025meta} & 0.117 & 0.087 & 0.122 & 0.077 & -0.243 & -0.104 & -0.190 & -0.187 \\ 
AES-PQ~\cite{tjandra2025meta} & 0.411 & 0.437 & 0.602 & 0.634 & 0.529 & 0.524 & 0.210 & 0.166 \\ 
UTMOS~\cite{baba2024utmosv2} & 0.026 & 0.013 & -0.010 & -0.024 & \textbf{0.610} & \textbf{0.591} & -0.178 & -0.154 \\
NISQA~\cite{mittag2021nisqa} & 0.104 & 0.119 & 0.158 & 0.183 & \underline{0.543} & \underline{0.549} & 0.243 & 0.212 \\
\midrule
Gemini-3-Pro$^+$~\cite{gemini}     & 0.175 & 0.165 & 0.532 & 0.537 & 0.141 & 0.169 & 0.159 & 0.154 \\
Gemini-2.5-Pro$^+$~\cite{comanici2025gemini25}   & 0.166 & 0.144 & 0.505 & 0.522 & 0.053 & 0.057 & -0.133 & -0.068 \\
Gemini-2.5-Flash$^+$~\cite{comanici2025gemini25} & 0.094 & 0.051 & 0.190 & 0.156 & 0.145 & 0.148 & 0.086 & 0.062 \\
Qwen3-Omni~\cite{xu2025qwen3}       & 0.231 & 0.186 & 0.341 & 0.311 & 0.113 & 0.104 & 0.052 & 0.065 \\
\midrule
\modelname\    & \underline{0.487} & \underline{0.484} & \textbf{0.642} & \underline{0.657} & 
0.496 & 0.461 & 
\underline{0.297} & \textbf{0.244} \\
\bottomrule
\end{tabular}
\begin{tablenotes}
    \footnotesize
    \item  $^+$ denotes models evaluated via API. All other models are inferred with the original code and weights.
    \item The best results are highlighted in \textbf{bold}, while the second-best results are \underline{underlined}.
\end{tablenotes}
\end{threeparttable}
\end{table*}

\subsection{Analysis of Prompt Robustness}

\begin{figure}[t]
    \centering
    \includegraphics[width=0.9\linewidth]{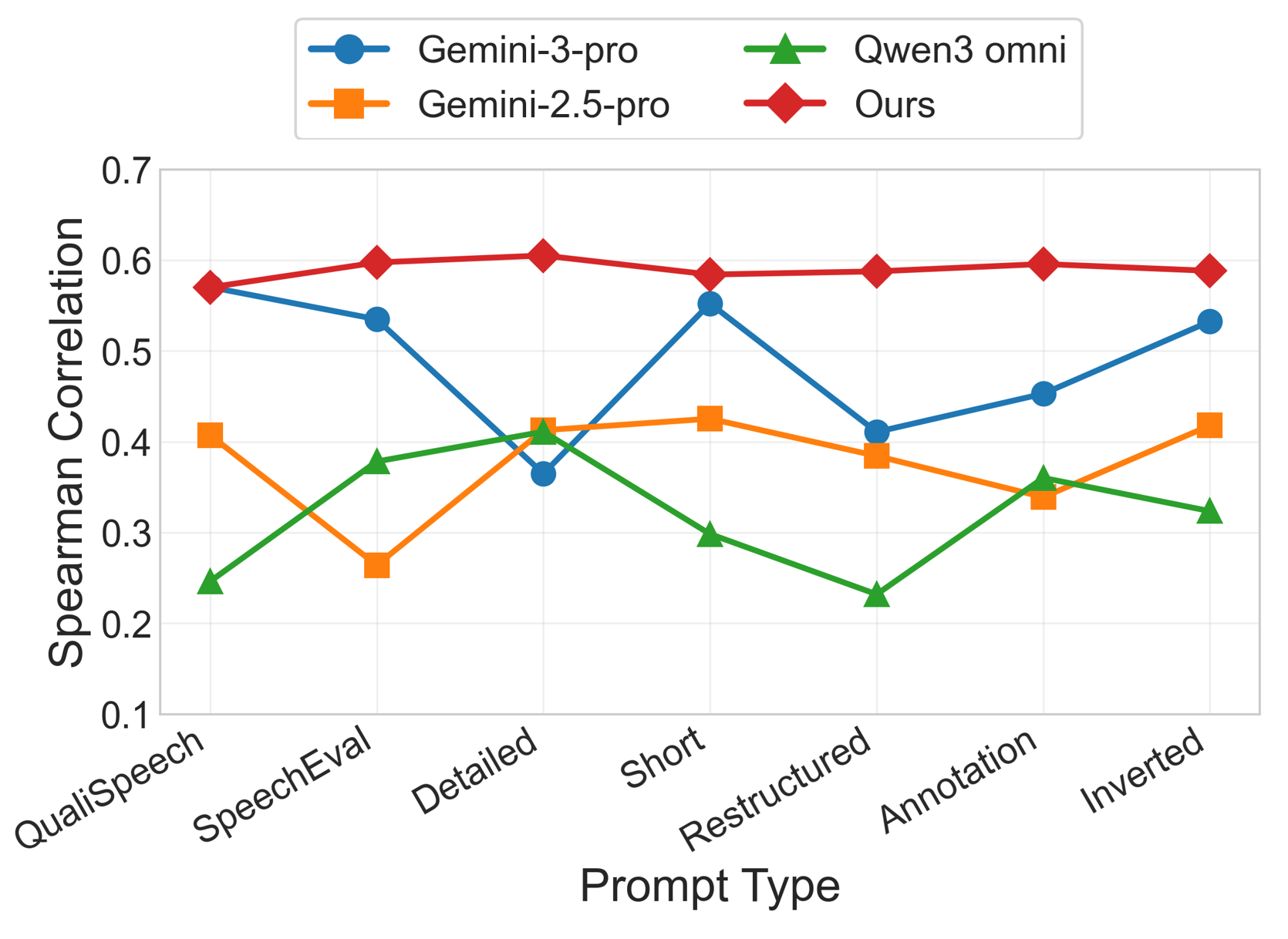}
    \caption{Cross-Model Spearman Correlation Comparison on Qualispeech Distortion Task with Various Task Description}
    \label{fig:distortion-prompt}
\end{figure}

\begin{figure}[t]
    \centering
    \includegraphics[width=0.9\linewidth]{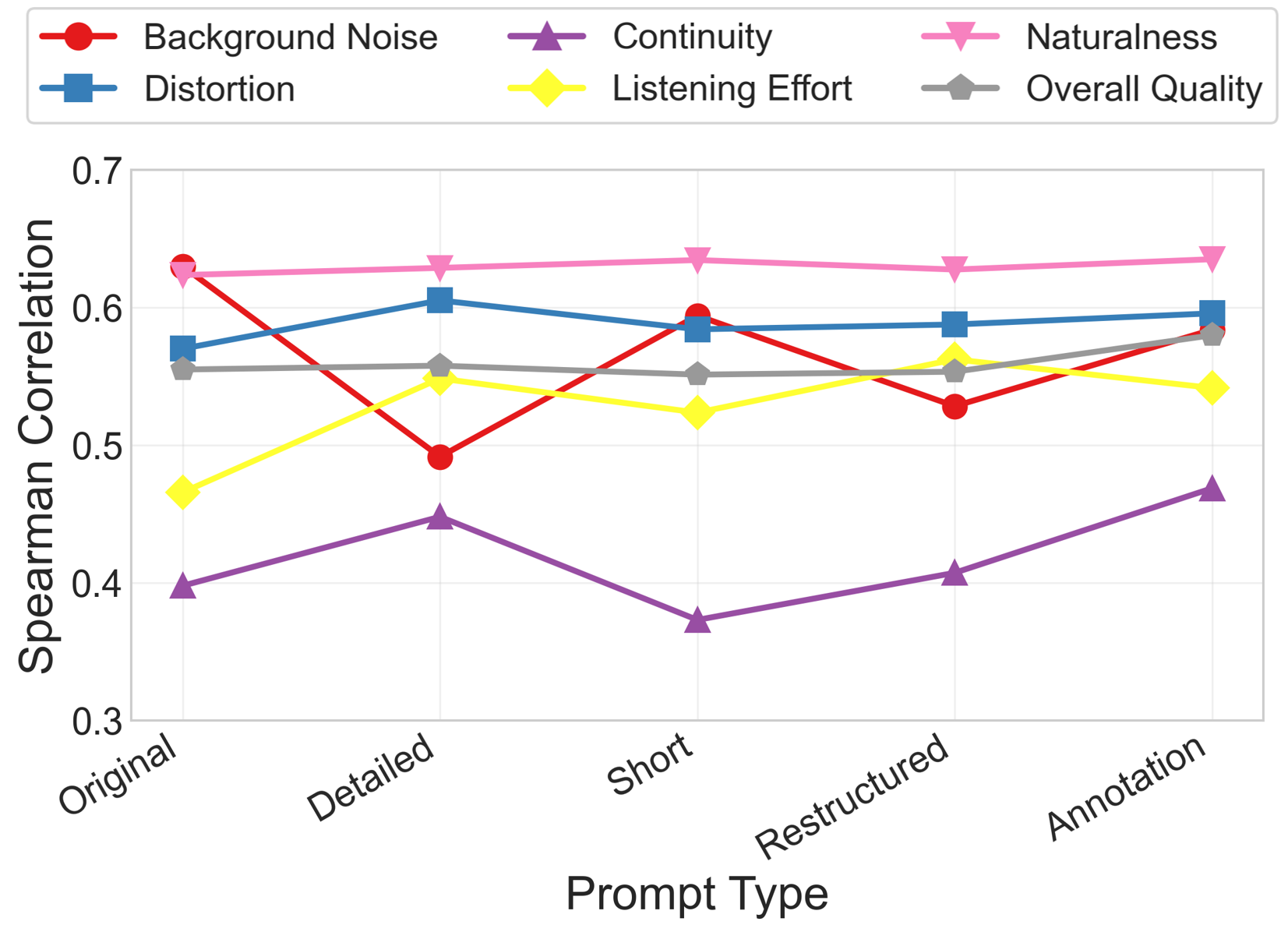}
    \caption{Cross-Metric Spearman Correlation Comparison of Our Model with Various Task Descriptions}
    \label{fig:ours-diff-prompt}
\end{figure}

LLMs are often highly sensitive to prompt engineering. For a truly robust evaluation framework, a model should adapt dynamically to varied instructions while maintaining performance consistency across semantically equivalent prompts. 

To assess the prompt robustness of our model, we performed a sensitivity analysis utilizing the QualiSpeech dataset. Unlike established "overall quality" metrics, this evaluation prioritizes distortion metrics, a non-standardized task that is subject to diverse technical interpretations. We generated a variety of prompt variants by adapting instructions from both QualiSpeech and SpeechEval. These variants include shortening descriptions (Short), altering grammatical formats (Restructured), adding granular details (Long), utilizing a more rigorous annotation protocol (Detailed) and inverting the metric scoring logic to ensure the model follows complex logical shifts (Inverted).

As illustrated in Figure \ref{fig:distortion-prompt}, our model demonstrates superior consistency across diverse prompt structures compared to baseline LLMs. While other models exhibited significant performance fluctuations even when the underlying task remained unchanged, our approach maintained a stable output, indicating high instructional resilience.

Furthermore, we evaluated the robustness of our model across a broad range of tasks and metrics, as illustrated in Figure \ref{fig:ours-diff-prompt}. Our model maintains highly stable performance across nearly all dimensions; however, the background Noisemetric is an exception, exhibiting increased sensitivity to specific prompt styles. Specifically, while performance remained consistent with original, short, and detailed instructions, the model struggled with long and restructured prompts, particularly those in passive voice, despite their semantic equivalence. This instability suggests a vulnerability to syntactic complexity, indicating that while the model captures the semantic intent of Noise evaluation, the increased ``attention cost" required to parse complex instructional structures may interfere with its ability to consistently quantify additive acoustic features. Despite this exception, the results across all other metrics remain robust, confirming the model’s overall instructional robustness.

\begin{table*}[htbp]
\centering
\caption{Ablation Study of Human-label data, Pseudo-label data and Proxy-task data}
\label{tab:data-source}
%\setlength{\tabcolsep}{2.8pt} % Slightly reduced to fit the extra columns
%\footnotesize
\begin{tabular}{c| ccc |cccc| cc| cc| ccc}
\toprule
\multirow{2}{*}{\textbf{System}} & \multicolumn{3}{c|}{\textbf{Data}} & \multicolumn{4}{c|}{\textbf{AES Speech}} & \multicolumn{2}{c|}{\textbf{QualiSpeech}} & \multicolumn{2}{c|}{\textbf{SpeechEval}} & \multicolumn{3}{c}{\textbf{AudioMOS}} \\
\cmidrule(lr){2-4} \cmidrule(lr){5-8} \cmidrule(lr){9-10} \cmidrule(lr){11-12} \cmidrule(lr){13-15}
~ & \textbf{Human} & \textbf{Pseudo} & \textbf{Proxy} & \textbf{CE} & \textbf{CU} & \textbf{PC} & \textbf{PQ} & \textbf{Dist} & \textbf{Ovrl} & \textbf{Dist} & \textbf{Ovrl} & \textbf{SynMOS} & \textbf{M-TA} & \textbf{M-Ovrl} \\
\midrule
\rowcolor[gray]{.9} 
\multicolumn{15}{c}{\textbf{Pearson Correlation (PCC $\uparrow$)}} \\
\midrule
S1 & \cmark & \cmark & \cmark & \underline{0.531} & \underline{0.594} & \textbf{0.601} & \textbf{0.707} & \underline{0.561} & 0.549 & \textbf{0.690} & \underline{0.662} & \textbf{0.496} & 0.487 & 0.642 \\
S2 & \cmark & \cmark & \xmark  & 0.512 & 0.580 & \underline{0.579} & 0.671 & 0.558 & \underline{0.577} & 0.662 & 0.608 & 0.461 & \textbf{0.537} & \textbf{0.694} \\
S3 & \cmark & \xmark  & \xmark  & -0.003 & 0.045 & -0.001 & 0.042 & \textbf{0.599} & \textbf{0.589} & \underline{0.685} & \textbf{0.684} & \underline{0.491} & 0.097 & 0.161 \\
S4 & \xmark  & \cmark & \xmark  & \textbf{0.550} & \textbf{0.595} & 0.425 & \underline{0.699} & -0.100 & 0.277 & 0.520 & 0.472 & 0.395 & \underline{0.493} & \underline{0.685} \\
\midrule
\rowcolor[gray]{.9} 
\multicolumn{15}{c}{\textbf{Spearman Correlation (SRCC $\uparrow$)}} \\
\midrule
S1 & \cmark & \cmark & \cmark & \underline{0.497} & \textbf{0.561} & \textbf{0.441} & \textbf{0.666} & \underline{0.570} & 0.555 & \textbf{0.685} & \underline{0.670} & \textbf{0.461} & 0.484 & 0.657 \\
S2 & \cmark & \cmark & \xmark  & 0.474 & 0.551 & \underline{0.418} & \underline{0.650} & 0.559 & \underline{0.571} & 0.645 & 0.625 & 0.454 & \textbf{0.535} & \textbf{0.712} \\
S3 & \cmark & \xmark  & \xmark  & 0.285 & 0.330 & -0.108 & 0.304 & \textbf{0.604} & \textbf{0.588} & \underline{0.673} & \textbf{0.697} & \underline{0.455} & 0.126 & 0.207 \\
S4 & \xmark  & \cmark & \xmark  & \textbf{0.514} & \underline{0.553} & 0.391 & 0.647 & -0.101 & 0.239 & 0.517 & 0.453 & 0.357 & \underline{0.500} & \underline{0.700} \\
\bottomrule
\end{tabular}
\end{table*}

\section{Ablation Study and Discussion}
\label{sec:ablation-limitation}

\subsection{Ablation study of Data Composition}

\subsubsection{Impact of Multi-Source Data}

As detailed in Section \ref{sec:data-aug}, our training framework utilizes multi-source, multi-task data. Table \ref{tab:data-source} presents an ablation study evaluating the impact of different data combinations. We observe that integrating all three sources, Human-labeled, Pseudo-Labeled, and Proxy data, yields the best performance across the majority of metrics.

Notably, the inclusion of proxy task data consistently enhances model performance by comparing S1 and S2. Moreover, the performance of S3 and S4 indicates that models trained on a single data type tend to overfit to that specific distribution, leading to poor generalization on unseen datasets. In contrast, diversifying data sources effectively broadens the model’s generalization capabilities for novel tasks and out-of-distribution evaluation.

\begin{table}[t]
\centering
\caption{Ablation Study of Template-based Augmentation (Step2) and LLM-Driven Paraphrasing (Step3)}
\label{tab:data_rewrite_strategy_pearson}
\setlength{\tabcolsep}{2pt}
\begin{tabular}{c| cc |cccc| ccc}
\toprule
\multirow{2}{*}{\textbf{Sys.}} & \multicolumn{2}{c|}{\textbf{Data}} & \multicolumn{4}{c|}{\textbf{AES Speech}} & \multicolumn{3}{c}{\textbf{AudioMOS}} \\
\cmidrule(lr){2-3} \cmidrule(lr){4-7} \cmidrule(lr){8-10}
~ & \textbf{Step2} & \textbf{Step3} & \textbf{CE} & \textbf{CU} & \textbf{PC} & \textbf{PQ} & \textbf{SynMOS} & \textbf{M-TA} & \textbf{M-Ovrl}\\
\midrule
\rowcolor[gray]{.9}  
\multicolumn{10}{c}{\textbf{Pearson Correlation (PCC $\uparrow$)}} \\
\midrule
D1 & \xmark & \xmark & \textbf{0.553} & \underline{0.594} & \textbf{0.587} & \textbf{0.718} & 0.267 & \underline{0.114} & \underline{0.148} \\
D2 & \cmark  & \xmark & -0.062 & -0.038 & 0.151 & 0.004 & \textbf{0.401} & -0.498 & -0.684\\
D3 & \cmark  & \cmark  & \underline{0.550} & \textbf{0.595} & \underline{0.425} & \underline{0.699} & \underline{0.395} & \textbf{0.493} & \textbf{0.685} \\
\midrule
\rowcolor[gray]{.9}  
\multicolumn{10}{c}{\textbf{Spearman Correlation (SRCC $\uparrow$)}} \\
\midrule
D1 & \xmark & \xmark & \textbf{0.521} & \textbf{0.556} & \textbf{0.405} & \textbf{0.657} & 0.299 & \underline{0.122} & \underline{0.140} \\
D2 & \cmark  & \xmark & 0.086 & -0.023 & 0.189 & -0.040 & \textbf{0.400} & -0.504 & -0.699\\
D3 & \cmark  & \cmark  & \underline{0.514} & \underline{0.553} & \underline{0.391} & \underline{0.647} & \underline{0.357} & \textbf{0.499} & \textbf{0.700}\\
\bottomrule
\end{tabular}
\end{table}

\subsubsection{Effectiveness of LLM-Driven Data Augmentation}

Table \ref{tab:data_rewrite_strategy_pearson} illustrates the effectiveness of our multi-calibration and LLM-driven multi-description augmentation strategies. In these experiments, all models were trained solely on AES pseudo-labeled data. It is important to note that the test set utilized the exact task descriptions from the fixed training setup (D1); consequently, these specific prompts were seen during training for the D1 but remained entirely unseen for D2 and D3.

Our results indicate that without data augmentation, model D1 overfits to specific task descriptions (e.g., CE, CU), effectively treating them as fixed categorical predictors rather than interpreting the underlying instructions. Simple templates D2 will fail to generalize to unseen test templates. In contrast, LLM-driven paraphrasing proves essential. By exposing model D3 to diverse, semantically equivalent instructions during training, the model learns to prioritize prompt intent over syntax. This approach not only yields performance on unseen prompts comparable to that of overfitted seen prompts but also demonstrates robust generalization capabilities.

\begin{table*}[htbp]
\centering
\caption{Ablation Study of Model Architecture on AES  Speech Sound and Music Datasets}
\label{tab:model_arch_aes}
% \setlength{\tabcolsep}{3pt}
% \footnotesize
\begin{tabular}{l |cccc |cccc |cccc}
\toprule
\multirow{2}{*}{\textbf{Model}} & \multicolumn{4}{c}{\textbf{Speech }} & \multicolumn{4}{c}{\textbf{Sound }} & \multicolumn{4}{c}{\textbf{Music }} \\
\cmidrule(lr){2-5} \cmidrule(lr){6-9} \cmidrule(lr){10-13}
& \textbf{CE} & \textbf{CU} & \textbf{PC} & \textbf{PQ} & \textbf{CE} & \textbf{CU} & \textbf{PC} & \textbf{PQ} & \textbf{CE} & \textbf{CU} & \textbf{PC} & \textbf{PQ} \\
\midrule
\rowcolor[gray]{.9} 
\multicolumn{13}{c}{\textbf{Pearson Correlation (PCC $\uparrow$)}} \\
\midrule
PEAF-base / GPT2        & 0.061 & -0.016 & 0.080 & -0.012 & 0.019 & -0.050 & 0.113 & -0.059 & 0.043 & 0.019 & 0.025 & -0.001 \\
PEAF-base / Llama1B & 0.515 & 0.563 & 0.587 & 0.700 & 0.557 & \underline{0.574} & \textbf{0.551} & 0.529 & 0.733 & 0.682 & 0.523 & 0.588 \\
PEAF-base / Llama3B & \textbf{0.531} & \textbf{0.594} & \textbf{0.601} & \textbf{0.707} & 0.542 & \textbf{0.618} & 0.495 & \textbf{0.569} & \underline{0.749} & \textbf{0.693} & \textbf{0.616} & \underline{0.669} \\
PEAF-small / Llama3B & \underline{0.515} & \underline{0.585} & 0.585 & \underline{0.707} & \textbf{0.591} & 0.570 & \underline{0.542} & 0.482 & 0.731 & 0.683 & \underline{0.598} & 0.664 \\
PEAF-large / Llama3B & 0.500 & 0.555 & \underline{0.590} & 0.698 & 0.554 & 0.568 & 0.539 & \underline{0.538} & \textbf{0.764} & \underline{0.692} & 0.589 & 0.664 \\
PEAV-base / Llama3B & 0.485 & 0.562 & 0.541 & 0.689 & \underline{0.583} & 0.571 & 0.531 & 0.534 & 0.743 & 0.679 & 0.523 & \textbf{0.671} \\
WavLM / Llama3B & 0.410 & 0.180 & 0.453 & 0.281 & -0.013 & 0.019 & -0.035 & 0.055 & 0.553 & 0.431 & 0.456 & 0.367 \\
\midrule
\rowcolor[gray]{.9} 
\multicolumn{13}{c}{\textbf{Spearman Correlation (SRCC $\uparrow$)}} \\
\midrule
PEAF-base / GPT2        & 0.067 & 0.034 & 0.052 & 0.031 & 0.039 & -0.070 & 0.161 & -0.088 & 0.035 & 0.017 & 0.016 & -0.009 \\
PEAF-base / Llama1B & 0.487 & 0.526 & \underline{0.427} & 0.658 & 0.535 & 0.537 & \textbf{0.555} & \underline{0.498} & 0.666 & \underline{0.638} & 0.466 & 0.582 \\
PEAF-base / Llama3B & \textbf{0.497} & \textbf{0.561} & \textbf{0.441} & \textbf{0.666} & 0.549 & \textbf{0.580} & 0.529 & \textbf{0.514} & 0.653 & 0.635 & \textbf{0.554} & 0.588 \\
PEAF-small / Llama3B  & \underline{0.494} & \underline{0.555} & 0.422 & \underline{0.665} & \textbf{0.564} & 0.542 & 0.546 & 0.447 & 0.653 & \textbf{0.638} & \underline{0.509} & 0.598 \\
PEAF-large / Llama3B  & 0.474 & 0.526 & 0.427 & 0.665 & 0.522 & 0.527 & 0.537 & 0.485 & \textbf{0.680} & 0.638 & 0.505 & \underline{0.602} \\
PEAV-base / Llama3B & 0.445 & 0.527 & 0.412 & 0.642 & \underline{0.554} & \underline{0.569} & \underline{0.551} & 0.481 & \underline{0.668} & 0.625 & 0.413 & \textbf{0.609} \\
WavLM / Llama3B & 0.429 & 0.283 & 0.387 & 0.454 & 0.324 & 0.307 & 0.482 & 0.393 & 0.514 & 0.424 & 0.386 & 0.411 \\
\bottomrule
\end{tabular}
\end{table*}

\subsection{Ablation Study of Model Architecture}

The proposed \modelname\ framework is designed to be agnostic to specific model architectures. Table \ref{tab:model_arch_aes} compares performance across different audio encoders and LLM backbones using the AES dataset.

\textbf{Audio Encoder:} Specialized audio encoders are vital for performance. However, we found that WavLM-base~\cite{chen2022wavlm}, which is optimized for speech, causes a performance degradation in general sound evaluation tasks. Moreover, regarding scale, the size of the audio encoder (Base vs. Large) appears less critical, as both configurations yield comparable results. The PE-AV audio encoder is less suitable for its specialized version PE-A-Frame~\cite{vyas2025pushing} as the latter is specialized trained on the audio event localization task so it has more detailed frame-level audio information.

\textbf{LLM Backbone:} Conversely, the scale of the LLM is a decisive factor. Small language models, such as GPT-2~\cite{radford2019languagegpt2}, lack the capacity to jointly model complex natural language instructions and audio features. Moving to larger scales, the 3B parameter model consistently outperforms the 1B version, demonstrating a superior ability to map diverse task descriptions to acoustic representations.

These trends remain consistent across the QualiSpeech and SpeechEval datasets, as further detailed in Table \ref{tab:model_arch_quali_speecheval}.

\begin{table*}[htbp]
\centering
\caption{Ablation Study of Model Architecture on Speech-only Datasets}
\label{tab:model_arch_quali_speecheval}
% \setlength{\tabcolsep}{3.5pt}
% \footnotesize % Using scriptsize to ensure 14 columns fit the page width
\begin{tabular}{l | cccccc | ccccccc}
\toprule
\multirow{2}{*}{\textbf{Model}} & \multicolumn{6}{c|}{\textbf{QualiSpeech}} & \multicolumn{7}{c}{\textbf{SpeechEval}} \\
\cmidrule(lr){2-7} \cmidrule(lr){8-14}
 & Noise& Dist. & Cont. & Listen. & Nat. & Ovrl. & Ovrl. & Int. & Dist. & Dyn. & Emo. & Art. & Subj. \\
\midrule
\rowcolor[gray]{.9} 
\multicolumn{14}{c}{\textbf{Pearson Correlation (PCC $\uparrow$)}} \\
\midrule
PEAF-base / GPT2 & 0.096 & 0.152 & 0.180 & 0.144 & 0.154 & 0.137 & 0.033 & 0.095 & 0.034 & 0.039 & -0.014 & -0.028 & 0.038 \\
PEAF-base / Llama1B  & 0.638 & 0.553 & 0.548 & 0.542 & 0.486 & \underline{0.582} & \underline{0.681} & \underline{0.673} & \underline{0.730} & 0.446 & \textbf{0.575} & \underline{0.502} & \underline{0.567} \\
PEAF-base / Llama3B  & 0.668 & 0.561 & 0.477 & 0.497 & \textbf{0.604} & 0.549 & 0.662 & 0.655 & 0.690 & \textbf{0.481} & \underline{0.564} & \textbf{0.509} & 0.534 \\
PEAF-small / Llama3B & \textbf{0.687} & \underline{0.567} & \textbf{0.616} & 0.505 & \underline{0.581} & 0.519 & \textbf{0.681} & 0.635 & 0.727 & 0.433 & 0.530 & 0.482 & 0.527 \\
PEAF-large / Llama3B & 0.650 & 0.566 & \underline{0.603} & 0.517 & 0.558 & 0.568 & 0.668 & 0.649 & \textbf{0.737} & 0.386 & 0.542 & 0.469 & 0.522 \\
PEAV-base / Llama3B & \underline{0.686} & 0.528 & 0.562 & \underline{0.566} & 0.553 & 0.560 & 0.646 & 0.672 & 0.686 & \underline{0.480} & 0.558 & 0.466 & 0.513 \\
WavLM / Llama3B    & 0.575 & \textbf{0.590} & 0.563 & \textbf{0.576} & 0.553 & \textbf{0.602} & 0.665 & \textbf{0.679} & 0.714 & 0.478 & 0.472 & 0.372 & \textbf{0.604} \\
\midrule
\rowcolor[gray]{.9} 
\multicolumn{14}{c}{\textbf{Spearman Correlation (SRCC $\uparrow$)}} \\
\midrule
PEAF-base / GPT2 & 0.068 & 0.110 & 0.197 & 0.107 & 0.118 & 0.090 & -0.046 & 0.015 & -0.039 & -0.030 & -0.103 & -0.085 & -0.001 \\
PEAF-base / Llama1B  & 0.591 & 0.553 & 0.549 & \underline{0.537} & 0.545 & \underline{0.581} & \underline{0.679} & 0.656 & \textbf{0.725} & 0.450 & \textbf{0.590} & \underline{0.499} & \underline{0.570} \\
PEAF-base / Llama3B  & 0.630 & 0.570 & 0.398 & 0.466 & \textbf{0.624} & 0.555 & 0.670 & 0.638 & 0.685 & 0.429 & \underline{0.567} & \textbf{0.506} & 0.542 \\
PEAF-small / Llama3B & \textbf{0.637} & \underline{0.572} & \textbf{0.618} & 0.474 & \underline{0.617} & 0.510 & \textbf{0.684} & 0.624 & 0.720 & 0.403 & 0.541 & 0.474 & 0.525 \\
PEAF-large / Llama3B & 0.603 & 0.568 & \underline{0.592} & 0.472 & 0.578 & 0.568 & 0.663 & 0.631 & \underline{0.723} & 0.356 & 0.545 & 0.460 & 0.509 \\
PEAV-base / Llama3B & \underline{0.628} & 0.538 & 0.568 & 0.534 & 0.578 & 0.560 & 0.641 & \underline{0.666} & 0.673 & \underline{0.460} & 0.566 & 0.468 & 0.502 \\
WavLM / Llama3B    & 0.531 & \textbf{0.612} & 0.556 & \textbf{0.571} & 0.606 & \textbf{0.605} & 0.676 & \textbf{0.669} & 0.709 & \textbf{0.466} & 0.490 & 0.401 & \textbf{0.619} \\
\bottomrule
\end{tabular}
\end{table*}

\subsection{Ablation Study of Training Steps}
\begin{figure}[t] % h:当前位置, t:顶部, b:底部, p:独立一页
    \centering % 使图片居中
\includegraphics[width=0.9\columnwidth]{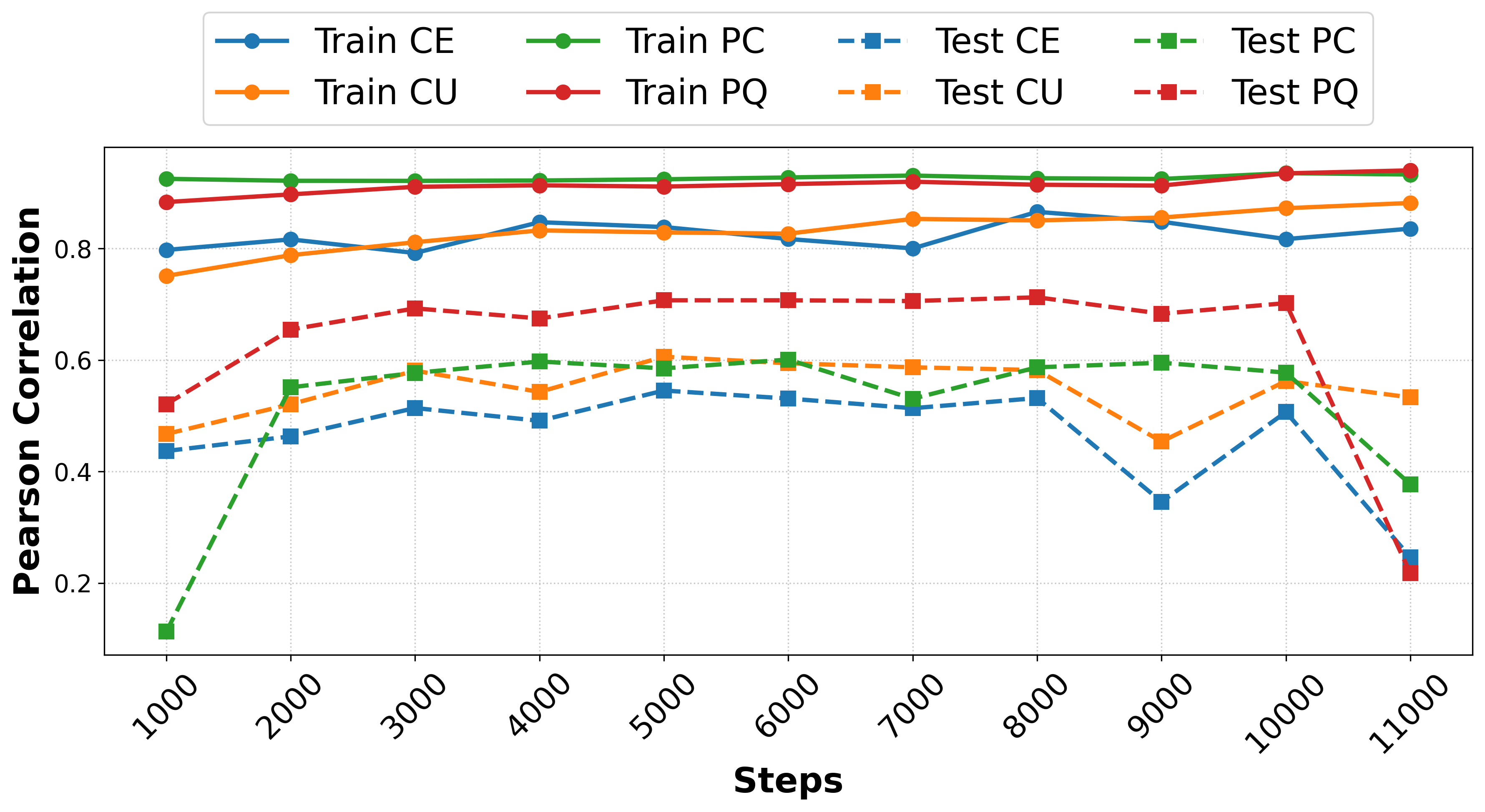} % 设置宽度并引入文件名
    \caption{Training and Inference Performance Comparison with the Different Training Steps} % 图片下方的文字说明
    \label{fig:train_infer_step} % 用于正文引用的标签
\end{figure}

We analyze the training dynamics of JASTIN by monitoring correlation metrics over 11,000 steps. As shown in Fig. \ref{fig:train_infer_step}, the model reaches peak performance at approximately 6,000 steps (about 1 epoch). Beyond this point, we observe a steady performance drop-off on the validation set, even though the training loss continues to decrease.

This phenomenon suggests that while the model continues to minimize loss on the training samples, it suffers from the catastrophic
forgetting of the LLM’s original abilities~\cite{lu2026desta2}. Specifically, when an LLM is tuned for too many iterations on a specialized regression task like audio assessment, it may lose the linguistic flexibility required to interpret diverse audio descriptions, leading to distributional bias toward the specific scoring patterns of the training set. To address this, we apply weight decay as a regularizer and implement an early-stopping mechanism to preserve the LLM's intrinsic reasoning capabilities while ensuring effective evaluation performance.

\subsection{Discussion of Failure Cases and Limitations}
Despite its overall robustness, our model exhibits performance bottlenecks when assessing temporally-sensitive metrics, specifically speech rate. As illustrated in Table \ref{tab:failure}, \modelname\ struggles to accurately categorize whether speech is appropriately paced, overly rapid, or too slow. 

This difficulty in capturing fine-grained temporal dynamics appears to be a systemic challenge for current MLLMs and our proposed method. General models like Gemini-2.5-Flash and Qwen3-Omni similarly exhibit low correlation on speed-related tasks. In contrast, advanced models like Gemini-3-Pro and Gemini-2.5-Pro achieve higher Pearson Correlation Coefficients. This improvement likely stems from their proficiency in audio-captioning and frame-level audio detection. Furthermore, because speed information is often orthogonal to perceived quality, a lower correlation may suggest that the model remains robust to variations in speaking rate. Overall, these results suggest that achieving precise rhythmic and durational analysis likely requires either increased model scaling or the integration of specialized temporal perception task.

Another failure mode is quality for specialized speech domains is still pretty low, for example, in the out-of-domain ASMR speech evaluation (as shown in Table \ref{tab:ood-performance}). In this context, almost all the LLMs, including our \modelname\ struggles to distinguish between high-fidelity aesthetic whispering and actual technical degradation. Because the model's internal speech-quality priors are rooted in voiced communication, the breathy, unvoiced nature of ASMR triggers a false-positive detection of artifacts or audio interference, reflecting a lack of domain-specific aesthetic sensitivity.

\begin{table}[t]
\begin{threeparttable}
\centering
\caption{Failure Case by analyzing Pearson and Spearman Correlation of Speaking Speed}
\label{tab:failure}
\begin{tabular}{l|c|c}
\toprule
\textbf{Model} & \textbf{Pearson Corr $\uparrow$} & \textbf{Spearman Corr $\uparrow$} \\
\midrule
AES-CE~\cite{tjandra2025meta} & 0.033 & 0.032 \\
AES-CU~\cite{tjandra2025meta} & 0.042 & 0.043 \\
AES-PC~\cite{tjandra2025meta} & -0.018 & -0.025 \\
AES-PQ~\cite{tjandra2025meta} & 0.026 & 0.027 \\
UTMOS~\cite{baba2024utmosv2} & 0.019 & 0.021 \\
NISQA~\cite{mittag2021nisqa} & 0.020 & 0.028 \\
\midrule
Gemini-3-Pro$^+$~\cite{gemini} & \underline{0.204} & \underline{0.225} \\
Gemini-2.5-Pro$^+$~\cite{comanici2025gemini25} & \textbf{0.247} & \textbf{0.250} \\
Gemini-2.5-Flash$^+$~\cite{comanici2025gemini25} & 0.011 & 0.140 \\
Qwen3-Omni~\cite{xu2025qwen3} & 0.083 & 0.087 \\
Qwen2-Audio~\cite{chu2024qwen2} & 0.042 & 0.031 \\
Audio-Flamingo3~\cite{goel2025audioflamingo} & -0.030 & -0.032 \\
\midrule
\modelname\  & 0.049 & 0.052 \\
\bottomrule
\end{tabular}
\begin{tablenotes}
    \footnotesize
    \item  $^+$ denotes models evaluated via API. All other models are inferred with the original code and weights.
    \item The best results are highlighted in \textbf{bold}, while the second-best results are \underline{underlined}.
\end{tablenotes}
\end{threeparttable}
\end{table}

Beyond temporal assessment, several promising avenues remain for future exploration. First, we plan to transition from single-audio assessment to multi-audio comparative paradigms. This will involve evaluating multiple samples simultaneously for relative ranking or integrating reference audios directly into the prompt to establish few-shot acoustic baselines. Second, we aim to move beyond scalar score regression by leveraging the generative capacity of the LLM backbone to provide interpretable diagnostic rationales. Future iterations will be trained to generate natural language critiques alongside numerical ratings, providing researchers with actionable feedback on specific acoustic artifacts.

\section{Conclusion}
In this work, we introduced \modelname, a novel framework that redefines automated audio evaluation by shifting from static, single-metric regression to a dynamic, instruction-driven paradigm. By bridging continuous, high-resolution acoustic features with the advanced cognitive capabilities of large language models and leveraging an interleaved conversational template and an LLM-guided data augmentation strategy, our proposed approach successfully enables context-dependent scoring and precisely aligns auditory inputs with diverse textual rubrics, closely mirroring the adaptability of human evaluators. Ultimately, \modelname\ establishes a more robust, flexible, and scalable foundation for the next generation of audio assessment.

\bibliographystyle{IEEEtran}
\bibliography{mybib}

\end{document}